\documentclass{optica-article}

\journal{opticajournal} % for journals or Optica Open

\articletype{Research Article}

\usepackage{lineno}
\usepackage{gensymb}
\usepackage{comment}
\usepackage{placeins}
\usepackage{float}

\begin{document}

\title{Versatile luminescence macroscope with dynamic illumination for photoactive systems}

\author{Ian Coghill,\authormark{1,5} Alienor Lahlou,\authormark{1,2} Andrea Lodetti,\authormark{3}
Shizue Matsubara,\authormark{3} Johann Boucle,\authormark{4} Thomas Le Saux,\authormark{1} and Ludovic Jullien\authormark{1,6}}

\address{\authormark{1}CPCV, Département de chimie, École normale supérieure, PSL University, Sorbonne Université, CNRS, 75005 Paris, France\\
\authormark{2}Sony Computer Science Laboratories, 75005 Paris, France\\
\authormark{3}Plant Sciences (IBG-2), Forschungszentrum Jülich GmbH, D-52425 Jülich, Germany\\
\authormark{4}Univ. Limoges, CNRS, XLIM, UMR 7252, F-87000 Limoges, France\\
\authormark{5}ian.coghill@ens.psl.eu\\
\authormark{6}ludovic.jullien@ens.psl.eu}
%% email address is required; see note below about the corresponding author designation

% use {asbstract*} to suppress the copyright line. Copyright information will be added in production

\begin{abstract*}
Luminescence imaging is invaluable for studying biological and material systems, particularly when advanced protocols that exploit temporal dynamics are employed. However, implementing such protocols often requires custom instrumentation, either modified commercial systems or fully bespoke setups, which poses a barrier for researchers without expertise in optics, electronics, or software. To address this, we present a versatile macroscopic fluorescence imaging system capable of supporting a wide range of protocols, and provide detailed build instructions along with open-source software to enable replication with minimal prior experience. We demonstrate its broad utility through applications to plants, reversibly photoswitchable fluorescent proteins, and optoelectronic devices.

\end{abstract*}

\section{Introduction}  %%%%%%%%%%%%% INTRODUCTION %%%%%%%%%%%%%%%%%%%%%%%%%%%%%%%%%%%%%%%%%%%%%%%%%%%%%%%%%%%
%%%%%%%%%%%%%%%%%%%%%%%%%%%%%%%%%%%%%%%%%%%%%%%%%%%%%%%%%%%%%%%%%%%%%%%%%%%%%%%%%%%%%%%%%%%%%%%%%%%%%%%%%%%%%%
%%%%%%%%%%%%%%%%%%%%%%%%%%%%%%%%%%%%%%%%%%%%%%%%%%%%%%%%%%%%%%%%%%%%%%%%%%%%%%%%%%%%%%%%%%%%%%%%%%%%%%%%%%%%%%
%%%%%%%%%%%%%%%%%%%%%%%%%%%%%%%%%%%%%%%%%%%%%%%%%%%%%%%%%%%%%%%%%%%%%%%%%%%%%%%%%%%%%%%%%%%%%%%%%%%%%%%%%%%%%%

Imaging is invaluable for observing and analyzing a variety of objects. In many applications, imaging instruments have been designed that go beyond traditional imaging, and additionally have components such as active light sources and filters to enable access to luminescence (which groups both fluorescence and phosphorescence) intensity, spectra, lifetime or polarization~\cite{valeur_molecular_2012}. Indeed, luminescence can produce extremely powerful observables, endowed with high specificity and versatility, enabling the state of observed systems to be unraveled.

\vspace{6pt}

The utility of luminescence-based imaging has expanded significantly with the advent of approaches that harness not only the intensity of luminescence but also its temporal dynamics~\cite{Querard2016}.  Multiple protocols that rely on tailored sequences of intensity-modulated light excitation have been engineered to operate reliably under demanding conditions, including high background autofluorescence, spectral overlap of luminophores, and interference from ambient light. In particular, they have found application in multiplexed microscopy imaging of spectrally overlapping reversibly photoswitchable fluorophores~\cite{richards_synchronously_2010,querard_resonant_2017,chouket_extra_2022,pellissier-tanon_correlation_2022,Pellissier2022a,valenta_per-pixel_2024,merceron_periodic_nodate}, macroscale fluorescence imaging under adverse optical conditions~\cite{zhang_macroscale_2018}, non-invasive mapping of tissue oxygenation using smart bandages~\cite{li_non-invasive_2014}, fluorescence imaging of plant physiological responses~\cite{kupper_analysis_2019, Nedbal2004}, and characterization of photovoltaic (PV) devices~\cite{sun_quantitative_2014,mandelis_method_2015,hu_ultrahigh-frequency_2019}. Given the extensive parameter space available in designing these protocols, there remains considerable opportunity to develop innovative illumination and acquisition strategies that enable rich, context-specific information to be extracted.

\vspace{6pt}

However, a major bottleneck in the development of such advanced luminescence-based imaging protocols is the lack of commercially available instrumentation capable of executing the required illumination and acquisition sequences. Researchers are frequently compelled to modify existing commercial systems or to construct entirely custom setups. Both approaches present significant challenges and are often inaccessible to researchers lacking interdisciplinary expertise in optics, optomechanics, electronics, and programming. Even when modification descriptions or custom designs are published, they are frequently documented with insufficient detail for straightforward replication.

\vspace{6pt}

Our group has recently focused on developing such protocols for plants, aiming to construct kinetic fingerprints that reflect their physiological state. In pursuing this work, we found no suitable commercially available macroscope system capable of imaging luminescence responses to user-defined sequences of light modulations across a wide range of frequencies, intensities, and wavelengths. There is a clear unmet need for an instrument with this level of flexibility, which could support not only plant-based research but also a broader range of applications. Such a tool would be valuable to both academic researchers and industrial users. Thus, in this manuscript, we present a highly versatile macroscopic fluorescence imaging instrument with sufficient information to enable replication at modest cost (<25 k€) by users with minimal expertise in instrumentation. The instrument offers: a sample area of up to approximately 1~cm~$\times$~1~cm; extensive flexibility in modulated illumination and acquisition sequences, with light modulations up to the 100 kHz range and camera frame rates up to 100 fps possible; multiple choices for excitation (5 in the range from 405 to 740 nm) and emission wavelengths (4 in the range from 540 to 740 nm); light intensities up to around 13,000 $\mu\mathrm{mol}\,\mathrm{m}^{-2}\,\mathrm{s}^{-1}$ [\textasciitilde
3800 $\mathrm{W\,m^{-2}}$]; the ability to use samples of various geometries, including small plants in pots. Once all of the commercial and 3D printed parts have been acquired, this build can take on the order of 2-4 full-time work weeks to complete. In fact, this instrument has already been duplicated for use in a plant research facility, and this was the approximate build time. This manuscript details: the capabilities of the instrument, its optical design, optomechanics, electronics and software; characterization, correction and calibration protocols. A full parts list, computer-aided design (CAD) files, detailed build instructions, as well as the software files, are provided. In providing these, in particular the CAD files, researchers with more advanced technical expertise will also be able to modify the design to their needs.

\vspace{6pt}

This manuscript also includes a series of demonstrations that highlight some of the instrument's capabilities. First, we apply a Pulse Amplitude Modulation (PAM)-like protocol, inspired by~\cite{Nedbal2004}, to image excised leaves and extract physiologically relevant parameters. The instrument is then used to image whole plants in pots, where sinusoidal light modulations are applied to capture frequency responses. We then demonstrate the application of some advanced imaging protocols (RIOM and HIOM~\cite{merceron_periodic_nodate}) on leaves, extracting kinetic fingerprints of their physiological state. Thereafter, we probe reversibly photoswitchable fluorescent proteins (RSFPs), present in droplets of solution, held between two microscope slides, using Speed-OPIOM~\cite{querard_resonant_2017} and RIOM. Finally, we apply a RIOM-like electroluminescence protocol to a solar cell and light emitting diode (LED). While these examples illustrate the instrument’s broad capabilities, it is designed to be highly versatile and is capable of supporting many other protocols.

\section{Instrument Design}     %%%%% INSTRUMENT DESIGN %%%%%%%%%%%%%%%%%%%%%%%%%%%%%%%%%%%%%%%%%%%%%%%
%%%%%%%%%%%%%%%%%%%%%%%%%%%%%%%%%%%%%%%%%%%%%%%%%%%%%%%%%%%%%%%%%%%%%%%%%%%%%%%%%%%%%%%%%%%%%%%%%%%%%%%
%%%%%%%%%%%%%%%%%%%%%%%%%%%%%%%%%%%%%%%%%%%%%%%%%%%%%%%%%%%%%%%%%%%%%%%%%%%%%%%%%%%%%%%%%%%%%%%%%%%%%%%
%%%%%%%%%%%%%%%%%%%%%%%%%%%%%%%%%%%%%%%%%%%%%%%%%%%%%%%%%%%%%%%%%%%%%%%%%%%%%%%%%%%%%%%%%%%%%%%%%%%%%%%

\subsection{Instrument Overview}\label{subsec:Instrument Overview} %%
\begin{figure}[htbp]
\centering
\includegraphics[width=12cm]{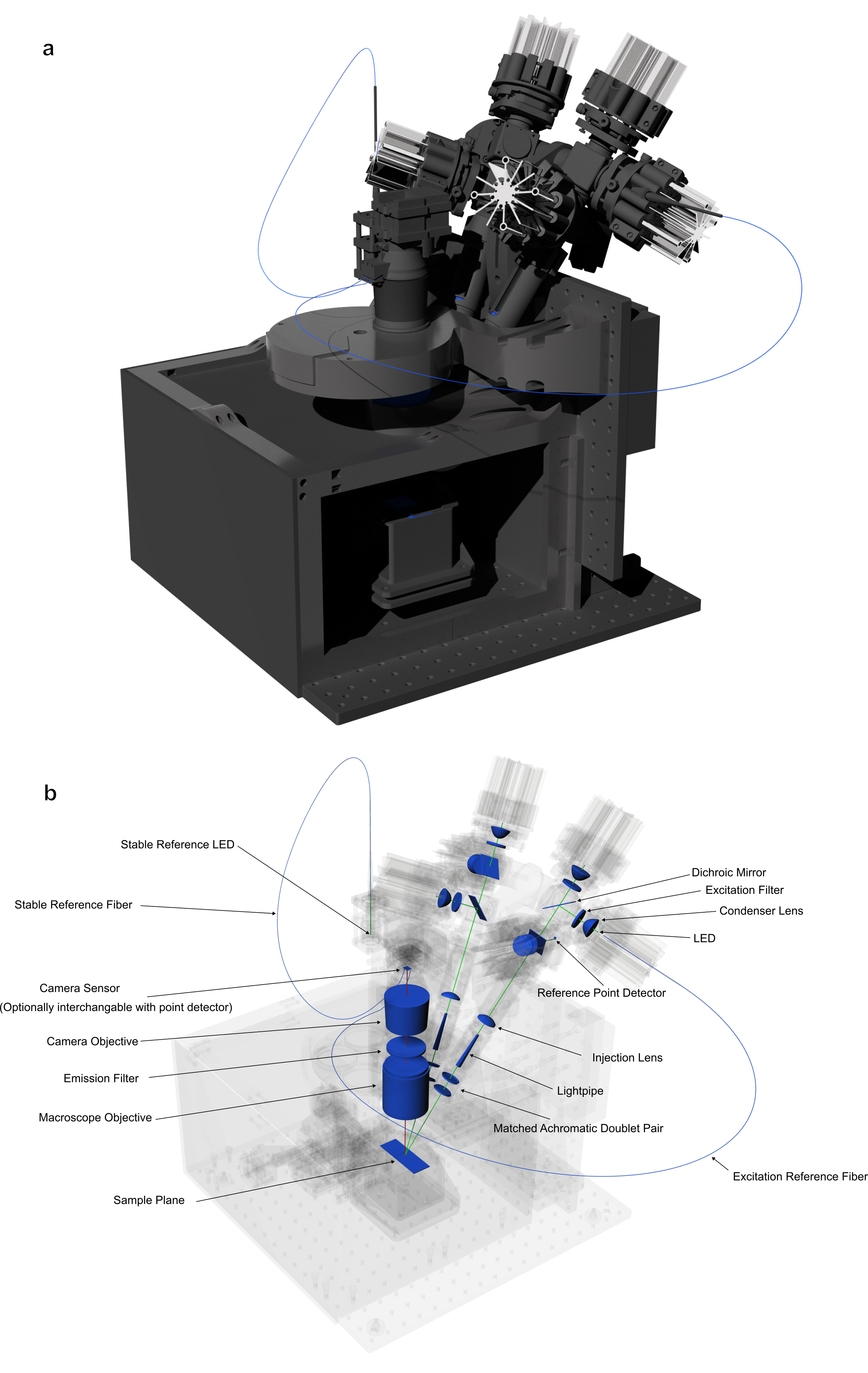}
\caption{\textbf{Normal (a) and ghosted (b) 3D CAD rendered images of the macroscope, showing the system design and optical components.} The instrument features independent illumination and imaging paths, making easy the use of different excitation and emission wavelength combinations. Two illumination arms, each offset by 40° from each other, and angled 30° from vertical, provide practically homogeneous illumination over a 10 × 10 mm$^2$ area (alternatively 5 × 5 mm$^2$, where higher intensities are needed). The imaging system captures a 21 × 13 mm$^2$ field with a high-resolution camera, which can be swapped for a point detector for high-speed measurements. The system supports synchronized illumination modulation and imaging. Many sample geometries and sizes can be accommodated, simplified by the large (70 mm) working distance. Some optical components which occur multiple times are not labeled more than once for clarity.} %\textcolor{red}{In b. interchang\textbf{e}able. Top/bottom vs Left/right.}}
\label{fig:FIG1}
\end{figure}

A CAD rendering of the instrument, along with a ghosted rendering that reveals its optical components, both created in Rhinoceros 3D (Robert McNeel \& Associates, Seattle, WA, US), are shown in Figure~\ref{fig:FIG1}. The side door has been omitted in order to show the location of the sample mount inside the dark box.

%%%%%%%%%%%%%%%%%%%%%%%%%%%%%%%%%%%%%%%%%%%%%%%%%% FIGURE 1
%%%%%%%%%%%%%%%%%%%%%%%%%%%%%%%%%%%%%%%%%%%%%%%%%%%%%%%%%%%
%%%%%%%%%%%%%%%%%%%%%%%%%%%%%%%%%%%%%%%%%%%%%%%%%%%%%%%%%%%%

\vspace{6pt}

The instrument contains an imaging path that looks straight down onto the sample, and two illumination arms which illuminate the sample from above but at an angle slightly offset from the imaging path. The decision to have the illumination paths independent of the imaging path, in contrast to an epi-type configuration, was made to simplify the use of different excitation and emission wavelength combinations, making working with a range of different luminophores more straightforward. Achieving the same flexibility with an epi-type configuration would typically require the availability of multiple dichroic mirrors, some of which might need to be custom-made. One drawback of our design is that, since the sample is illuminated from an angle, a slight gradient in light intensity is introduced. Yet, as shown in the optical design and calibration sections of this manuscript, it is minimal and unlikely to impact most applications. It is important to note that the angled illumination may also affect penetration depth, reflection, and fluorescence profiles, particularly in scattering media, so users should take this into account when working with such samples.

\vspace{6pt}

The two illumination arms are each offset from each other by 40{\degree}, according to the imaging system's optical axis, and project light onto the sample at an angle of 30{\degree} from vertical. Each arm houses three LED sources at distinct wavelengths, and delivers light to the same, approximately $10 \times 10~\text{mm}^2$, area. Specifically, the right arm is equipped with sources at 405~nm (20~nm bandwidth), 470~nm (40~nm bandwidth), and 645~nm (30~nm bandwidth), while the left arm features sources at 535~nm (30~nm bandwidth), 645~nm (30~nm bandwidth), and 740~nm (40~nm bandwidth). The spectral information of the sources, excitation filters and dichroic mirrors can be found via their manufacturers. This dual-arm configuration enables the simultaneous use of two identical wavelength sources, exemplified by the 645~nm channels, which facilitates decoupling of light sources and an additive increase in light intensity at a particular wavelength. These illumination arms can be reconfigured relatively easily, by swapping in different dichroic mirrors, excitation filters and LEDs, to accommodate different experimental protocols. In fact, in two of the demonstrations made in this work, those of the PAM-like and HIOM protocols, the arms were reconfigured such that each arm featured only a single 470~nm light source.

\vspace{6pt}

The instrument has two different ways of controlling the light sources, here high power LEDs, each with different advantages and limitations. In one way, an LED driver (whose output current follows a supplied modulation voltage signal) is used to power them. In the other, they are directly driven using the output of a waveform generator. The LED driver is able to drive the LEDs with much higher current levels, typically 1 A, but it comes at the cost of a progressively changing waveform amplitude and shape towards higher frequencies. This is demonstrated for both sinusoidal and square waveforms in Figure S92, where at high drive currents, the shape of the square waveform is particularly troublesome. The use of a driver was therefore not possible for us in the development of our RIOM protocol, and we often did not use the LED driver for experiments requiring modulation faster than 1 kHz. Here the waveform generator was useful. It gives a consistent waveform up to high frequency (Figure S92), but at the cost of a lower maximum light intensity since it can only supply roughly 0.2-0.3 A. It can generate electrical signals up to 80 MHz, yet the LEDs can only be modulated until around 100 kHz without too much distortion, since beyond this they begin to not be able to respond fast enough. Regarding light intensities, the maximum values we achieved by both means, using an illumination area of $10 \times 10~\text{mm}^2$, are presented in Table~\ref{tab:irradianceLevels}. Even higher light intensities can be obtained by reducing the illumination area to $5 \times 5~\text{mm}^2$, which is possible by making a small modification: flipping the lightpipe and changing its injection lens, as described in Section 4 of the Supporting Information. These are also provided in Table~\ref{tab:irradianceLevels}. The intensities here were determined using a power meter, in the manner detailed in Subsection~\ref{subsec:Calibration of irradiance}. The intensities provided in the table reflect those achieved while the light sources and dichroic mirrors are arranged as shown at the bottom left of Figure S34, yet it can be possible to achieve even higher light intensities at individual wavelengths by removing all dichroic mirrors and mounting only a single light source on a particular illumination arm, avoiding optical losses due to mirror inefficiency. The LED driver can drive four LEDs simultaneously. The DAQ device can supply only two analog modulation signals; the remaining channels are limited to digital (i.e., on/off) control. However, if more than two analog signals are required, the waveform generator can be used to provide two additional ones. When driving LEDs directly with the waveform generator, only two LEDs can be driven simultaneously.

\begin{table}[t!]
    \caption{Maximum intensities accessible on the instrument for two control configurations: LED driver and waveform generator. The square illumination area can be set to either $10 \times 10$ mm\textsuperscript{2} or $5 \times 5$ mm\textsuperscript{2}, modifying the maximum achievable intensity levels.}
    \label{tab:irradianceLevels}
    \centering
    \begin{tabular}{ |c|c|>{\centering\arraybackslash}p{3cm}|>{\centering\arraybackslash}p{3cm}| }
        \hline
        \multirow{2}{*}{$\lambda$ (nm)} & \multirow{2}{*}{Device} & \multicolumn{2}{c|}{Maximum Intensity ($\mu \mathrm{mol} \, \mathrm{m}^{-2} \mathrm{s}^{-1}$) [$\mathrm{W\,m^{-2}}$]} \\
        \cline{3-4}
        & & $10 \times 10~\text{mm}^2$ & $5 \times 5~\text{mm}^2$ \\
        \hline
        \multirow{2}{*}{405} & LED driver & 4400 [1300] & 13500 [3990] \\
                             & Waveform generator & 790 [233] & 2500 [739] \\
        \hline
        \multirow{2}{*}{470} & LED driver & 5200 [1324] & 13800 [3515] \\
                             & Waveform generator & 1290 [328] & 3200 [815] \\
        \hline
        \multirow{2}{*}{535} & LED driver & 2600 [482] & 8700 [1615] \\
                             & Waveform generator & 390 [72] & 1300 [241] \\
        \hline
        \multirow{2}{*}{645} & LED driver & 2000 [447] & 6600 [1477] \\
                             & Waveform generator & 600 [134] & 2000 [447] \\
        \hline
        \multirow{2}{*}{740} & LED driver & 1900 [307] & 3800 [615] \\
                             & Waveform generator & 340 [55] & 700 [113] \\
        \hline
    \end{tabular}
\end{table}

\vspace{6pt}

Regarding imaging, the field of view is \(21 \times 13~\text{mm}^2\), at a working distance of 70 mm from the macroscope objective, a configuration that facilitates compatibility with various sample types, including Petri dishes, microscope slides, and small plants in pots. Despite the relatively large working distance, the objective’s substantial diameter enables a satisfactory numerical aperture (NA), providing effective light collection. This however comes at the cost of reduced depth of field, limiting its suitability for samples with greater depth. In such cases, the camera objective can be stopped down to reduce the NA, increasing the depth of field, at the cost of reduced light collection efficiency. The imaging path features emission filters for luminescence detection at 540~nm (50~nm bandwidth), 632~nm (60~nm bandwidth), 690~nm (50~nm bandwidth), and 740~nm (40~nm bandwidth). The spectral information of these filters can be found via their manufacturers. Images are captured by a grayscale global-shutter CMOS camera, operable at either 12- or 8-bit, with a resolution of \(1936 \times 1216~\text{pixels}\). A global shutter, rather than a rolling shutter, was favored since the instrument was designed for protocols involving dynamic signals, where the timing relative to the illumination sequence must be known for all pixels. Although a rolling shutter camera can be employed, it adds complexity for such protocols. While the camera’s specifications list a maximum frame rate of 166 fps, we have reliably operated it up to 100 fps using external triggering. Further details on the camera, including its spectral response, can be found in the datasheet available from the manufacturer. If one wishes to change the camera on this setup to one of higher performance, it would not be overly complex to do so, but would require changes to the supplied software scripts. For applications not requiring imaging, the camera can be replaced with a point detector, allowing sampling rates of up to approximately 1 MHz with the current setup. Faster DAQ devices could enable sampling rates beyond this.

\newpage

The timing and synchronicity of camera/point detector acquisition and the light sources is ensured using a DAQ device, ensuring a fixed phase relationship between illumination and image acquisition. As a potentially useful addition, an optical fiber has been added to collect stray light from one LED (illustrated in Figure S99) being piloted to one edge of the camera sensor (outside the sample imaging region) providing direct access to the phase of the illumination waveform within the recorded images, facilitating precise phase comparison with the luminescence signal. This feature additionally enables verification that the illumination levels remain stable and undistorted, and do not drift due to LED heating, though we have confirmed that the LEDs on the system have a very stable output over hours, even at maximum drive current (Figure S100). Beyond potential LED heating, other instrumental artifacts may also influence experimental results. A notable example is camera heating, which can be a factor here since the camera used is not temperature-regulated, unlike expensive scientific cameras. Consequently, a second optical fiber was added to direct light from an LED light source, driven at constant low power to be as stable as possible, to another edge of the image sensor, enabling camera signal level changes caused by camera heating to be detected and accounted for. An example of each of the reference signals mentioned here, collected during one of the demonstration experiments, is shown in Figures~\ref{fig:FIG5}e,f. If instead it is desired to have reference signals for two independently piloted light sources, the fiber connected to the stable LED source can be instead connected to one of the piloted sources. The capture of the illumination reference signals is limited to the acquisition frequency of the camera. In cases where recording of the illumination waveforms needs to be done at higher frequency, a point detector can be added to either illumination arm, allowing it to be sampled at up to 1 MHz. Here the output of only one light source can be recorded at a time. As an example, one has been added to the right illumination arm in Figure~\ref{fig:FIG1}. Figures S39 and S40 show this from a better viewpoint.

\vspace{6pt}

The sample stage allows fine movement in the x, y, and z directions, either by using the stepper motors or by hand, for positioning and focusing prior to imaging. When operated using the motors, the movement resolution is on the order of 1~$\mu$m. The sample platform features a small LED light source in close proximity to the sample location (see Figures S12-17) for illuminating the sample while it is being positioned and focused. The LED emits light at 850 nm, providing visibility while avoiding the excitation bands of many luminophores. A temperature and humidity sensor has also been attached to the sample platform, in close proximity to the sample location (see Figures S12-17), enabling real-time environmental monitoring during experiments. To minimize interference from external light, the sample is enclosed within a black box. This enclosure features removable doors for easy access.

\vspace{6pt}

Experiments are written as Python scripts, providing a lot of flexibility and enabling the execution of a wide range of experimental protocols. Additionally, a framework has been implemented to automate the storage of data and metadata in an organized database, ensuring that all necessary information is available for reproducing past experiments.

\subsection{Optical Design} \label{subsec:Optical Design}

In this subsection, the design of the illumination and imaging paths is mainly dealt with, with only some comments on the other optical parts since their design was very straightforward. The optical components can be seen in position in the ghosted CAD view in Figure~\ref{fig:FIG1}.

\vspace{6pt}

The design of the imaging path involved simply the selection of high-quality commercially available objectives (ensuring high correction for aberrations) which give a magnification of around 0.5 (giving a field of view slightly larger than the illuminated area), a flat field, and high NA (to optimize light collection efficiency) with a working distance that accommodates use with a variety of different sample types (including plants). The optics selected were a 1X plan achromatic lens (\textit{f} = 100 mm) and a camera objective (\textit{f} = 50 mm) with adjustable f-number from f/1.8 to f/22. Regarding the camera, it was selected based on a combination of cost (since our work is often carried out with low-cost applications in mind), frame rate, and pixel size. Since the instrument was designed to measure often weak fluorescence signals and their dynamics, frame rate and pixel size were prioritized over spatial resolution. The spatial demands of the applications targeted are modest. As an example, for imaging leaves it is typically sufficient to resolve major features such as leaf veins. In \textit{Arabidopsis} seedlings, which represent the smallest samples we aim to probe, veins are approximately $50\,\upmu\mathrm{m}$ wide. Given the system’s optical magnification of $0.5\times$ and a sensor pixel size of $5.86\,\upmu\mathrm{m}$, this corresponds to an effective sampling of approximately $11.7\,\upmu\mathrm{m}$ per pixel in object space. These smallest veins therefore span about 4 to 5 pixels, which is sufficient for reliable discrimination. In fully developed leaves or larger plant species, vein diameters are considerably larger, making resolution even less demanding. To improve performance under low-light conditions, a sensor with relatively large pixels was chosen to enhance sensitivity by increasing photon collection per pixel. A sufficiently high frame rate was also important to provide good temporal resolution, allowing sufficient samples per period for dynamic fluorescence signals with frequency components up to around $20\,\mathrm{Hz}$.

\vspace{6pt}

Theoretical calculations (detailed in Section 7 of the Supporting Information) were performed based on the data provided by the manufacturers of the chosen lenses (too scarce to run an optical simulation), in order to give a rough indication of the NA, depth of field, and resolution which could be expected from this configuration. Those predict that the system would have an adjustable NA ranging from 0.016 to 0.19, and accordingly a total depth of field ranging from 0.135 to 3.5 mm (at 540 nm), allowing for work with samples of different geometries/thicknesses, including plants with leaves at different heights (Figure~\ref{fig:FIG5}e), and seeds (Figure S103). Based on the NA, the resolution (in the form of the Rayleigh resolution) was predicted to range from 1.7 to 20.3 $\mu$m at 540~nm. Here the resolution will be limited by the spatial sampling resolution, determined to be 23.5 $\mu$m according to the Nyquist criterion. Given this, the measurement of the point spread function (PSF) would be undersampled and thus not possible (as evidenced by the image taken of 200~nm beads loaded with Nile Red given in Figure S97). The dependence of the NA with f-number, and depth of field and resolution with f-number and wavelength, is provided in Figures S93, S94, and S95, respectively. In some applications, field flatness, of which both objectives are corrected for, can be an important aspect. In order to evaluate this, 1 $\mu$m beads loaded with Nile Red dried onto the surface of a glass slide were imaged at the highest NA setting, and upon assessing zoomed-in regions across the $10 \times 10~\text{mm}^2$ zone (Figure S98), it can be seen that focus is well maintained.

\vspace{6pt}

In using optics with a large diameter, it meant using large emission filters, 50 mm in diameter, in order to avoid cutting off light. These are positioned in the infinity space between the two objectives. Considering the emission filters, NA, sensor quantum efficiency, and assumed transmission values of the optics, the overall collection efficiency of the instrument was estimated to range from 0.004 to 0.6 \% (depending on the f-number selected) at 540~nm. The dependence of estimated overall collection efficiency with f-number for each emission wavelength is provided in Figure S96.

\vspace{6pt}

In the case where the point detector is used instead of the camera, a condenser lens (\textit{f} = 20 mm) is placed between it and the camera objective, reducing the image size given the $3 \times 3~\text{mm}^2$ size of the detection element. 

\vspace{6pt}
 
Regarding the optical fibers used for the reference signals on the camera, an optical fiber with a relatively small diameter ($200~\mu\mathrm{m}$) and low NA (0.22) was selected in order to have a low exit light cone angle, limiting the spread of light across the sensor.

\newpage

The design of the illumination arms was supported by ray tracing-based simulations made in OpticStudio 18.9 (Zemax LLC, Kirkland, WA, US) since to achieve highly homogeneous intensity over a relatively large area is not trivial. Out of the many strategies which exist (e.g., using diffusers, lightpipes or microlens arrays), here a lightpipe-based design was selected for its simplicity, high optical throughput, and ability to produce a well-defined square illumination zone that aligns well with the rectangular field of view. As can be seen in Figure~\ref{fig:FIG1}b, the optical configuration for each arm features high NA condensers (\textit{f} = 16 mm) to collect and collimate light from each individual LED ($\sim\!1 \times 1\,\text{mm}^2$); and bandpass filters to cut unwanted parts of the sources' emission spectra. Thereafter, dichroic mirrors are used to combine the light paths into a common light path before a condenser lens (\textit{f} = 20 mm) is used to inject that light into the small end of a lightpipe ($2.5 \times 2.5~\text{mm}^2$ entrance, $5 \times 5~\text{mm}^2$ exit, 50 mm long) to homogenize the light through multiple internal reflections, generating a strongly homogeneous square-shaped output at its exit face. Then, the exit face is simply relayed to the sample plane using a matched achromatic doublet pair ($f_1 = 40~\text{mm},\ f_2 = 100~\text{mm}$). 

\vspace{6pt}

%%%%% FIGURE 2
\begin{figure}[b!]
\centering
\includegraphics[width=13.2cm]{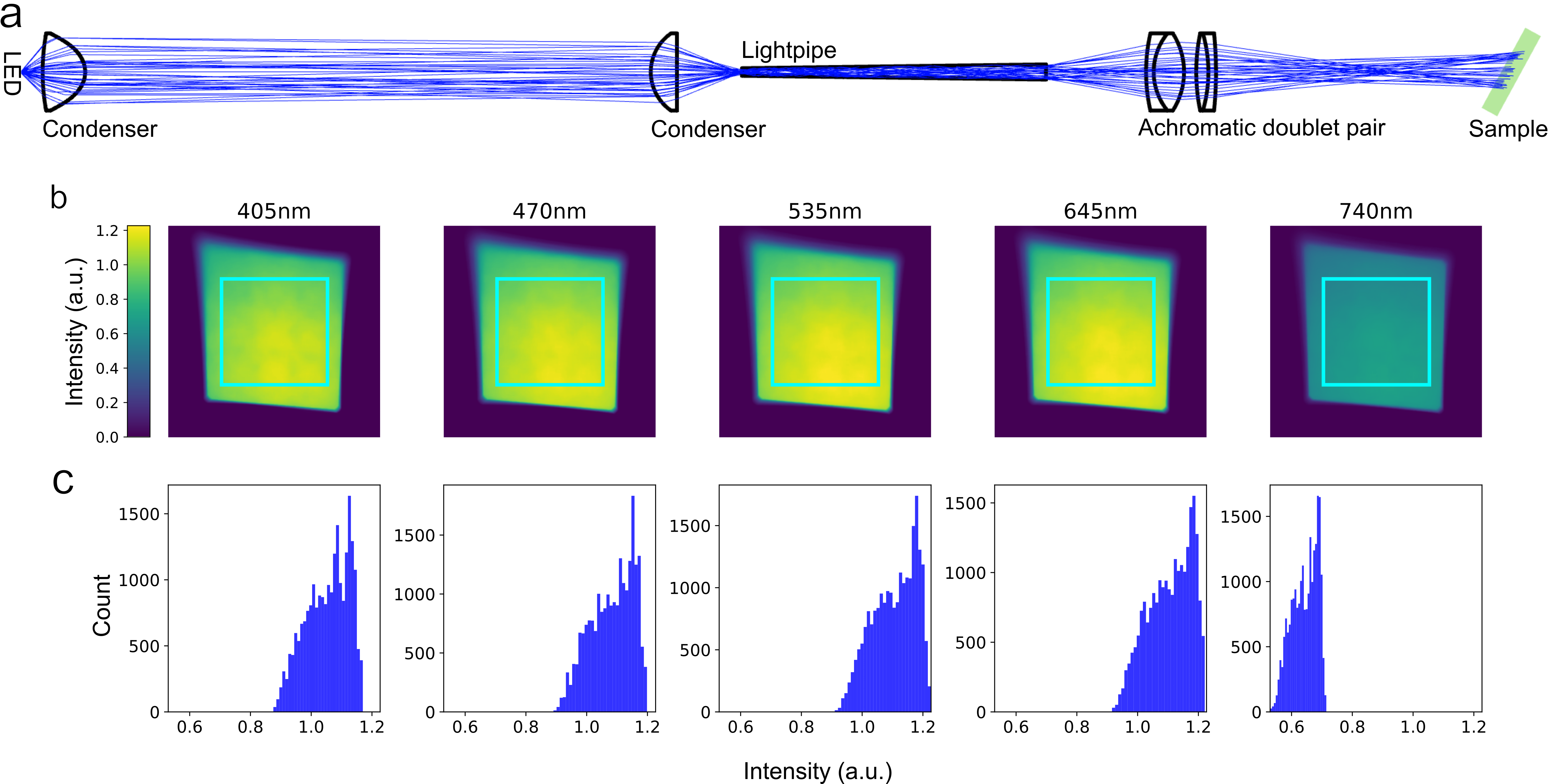}
\caption{\textbf{Optical simulation of the illumination path.} a) Optical model implemented in OpticStudio. It features a high-NA condenser lens to collect and collimate the light emitted by the LED; a condenser lens to inject this light into a lightpipe; and an achromatic doublet pair to relay the lightpipe's exit to the sample plane. Excitation filters and dichroic mirrors were not implemented in the model. b,c) Maps and histograms of the optical power over the sample plane for each wavelength source available on the instrument, showing a high degree of homogeneity.}
\label{fig:FIG2}
\end{figure}

The model implemented in OpticStudio for the $10 \times 10~\text{mm}^2$ configuration is shown in Figure~\ref{fig:FIG2}a. The model implemented features only a single LED, and no excitation filters or dichroic mirrors, given that their impact on the optical performance is negligible. Each wavelength was tested simply by changing the source wavelength. Importantly, the sample plane was tilted in accordance with the tilt of the illumination arms. Figures~\ref{fig:FIG2}b,c show the maps of intensity obtained (after removal of simulation noise), and the histograms of the intensity values within the central $10 \times 10~\text{mm}^2$ zone. The corresponding coefficients of variation computed range from 6.47-6.58~\%. The same analysis was performed for the $5 \times 5~\text{mm}^2$ configuration, giving coefficients of variation of between 2.04 and 3.16~\%. The corresponding figures, as well as the full details on the simulation and analysis steps performed, are provided in Section 3 of the Supporting Information.

\subsection{Optomechanical Design}

Given the large number of components, the full optomechanical design is not detailed here. The reader is referred to the build instructions in the Supporting Information, where all components are visible and their functions can be understood, as well as to the complete CAD design file provided with this manuscript. We note that many of the optomechanical parts were obtained commercially, but several were custom-fabricated using 3D printing. The use of many 3D-printed parts makes it straightforward to modify the design of different components when needed, and changes to the instrument can be implemented quickly.

\subsection{Electronics}

\begin{figure}[t!]
\centering
\includegraphics[width=11cm]{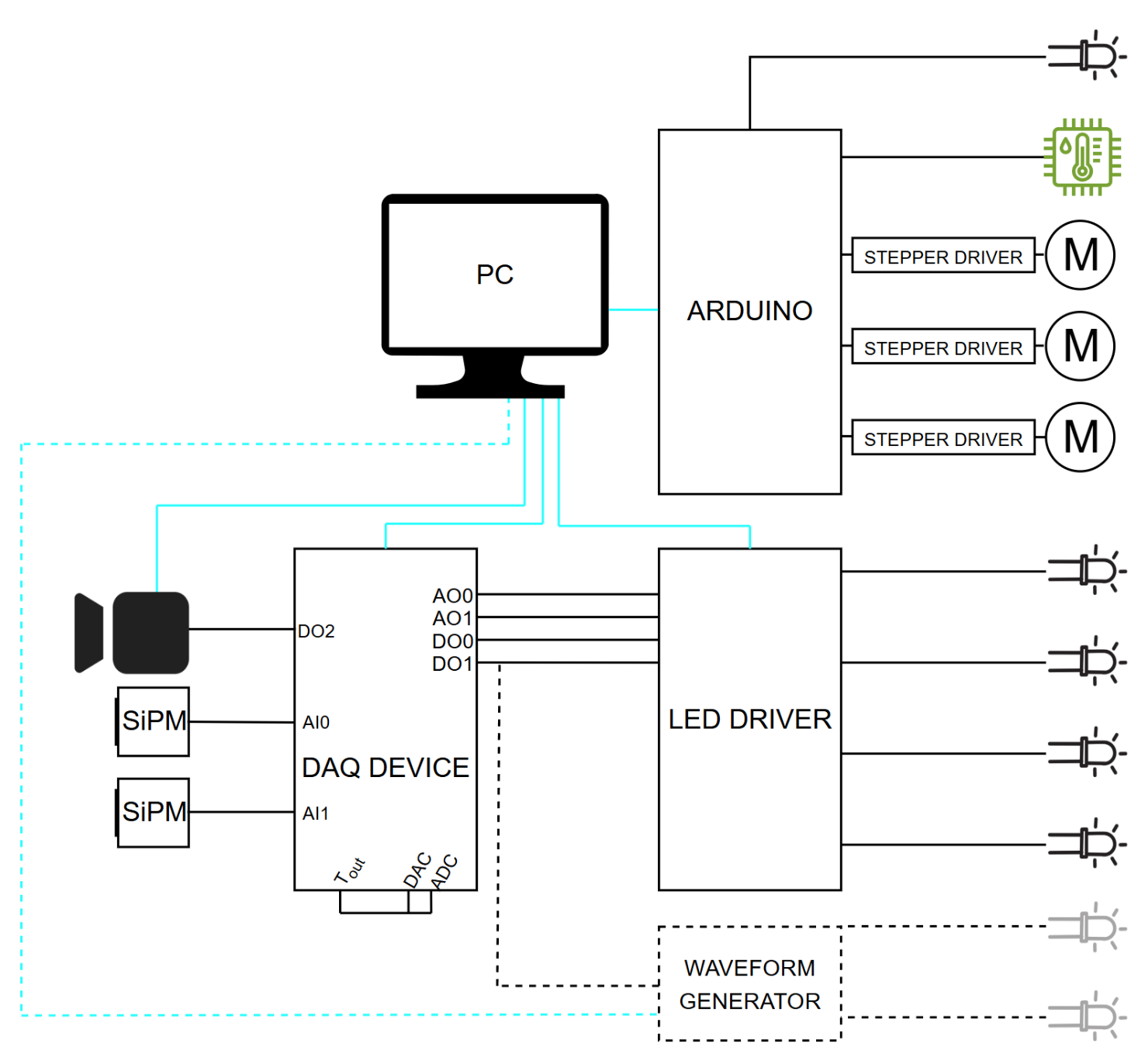}
\caption{\textbf{Schematic of the electrical setup.} The DAQ device synchronizes all of the electro-optical components. It controls LED modulation or triggers the waveform generator; acquires point detector signals; and triggers the camera. In the case of driving the LEDs with the LED driver, it outputs current levels which are proportional to the voltage levels sent by the DAQ device. In the case of the waveform generator, more suitable for higher modulation frequencies, the LEDs are driven directly. An Arduino Uno microcontroller board controls the 850 nm guide light LED, temperature and humidity sensor, and x,y,z stepper motors (via drivers). Power and ground lines are omitted for clarity. Cyan lines indicate connections to the PC.}
\label{fig:FIG3}
\end{figure}

A basic diagram of the instrument's electrical components is provided in Figure~\ref{fig:FIG3}. For clarity, power supply and ground lines are omitted. The diagram also shows, as dotted lines, the addition to the scheme when it is desired to use the waveform generator to directly drive LEDs. As can be seen in the diagram, the Arduino, DAQ device, LED driver, camera and waveform generator are all connected to the PC via USB cables, allowing for them to be communicated with. 

%%%%% FIGURE 3

\vspace{6pt}

The LED driver drives LEDs with currents proportional to the voltages applied to its modulation inputs: 100 mA for every 1 V supplied, up to a maximum of 1 A. Two of the modulation inputs are fed by the DAQ device's two analog outputs, having a range of 0 to 10 V. The remaining two modulation signals are supplied by the device's digital outputs. The digital outputs can provide either 0 or 3.3 V. Although not implemented here, an operational amplifier circuit could be used to boost the digital output voltage to 10 V, in order to be able to reach the maximum modulation level of the LED driver. When the waveform generator is used instead of the LED driver, the drive signals to be output are pre-programmed, and one of the DAQ device's digital outputs is used to trigger its start. The DAQ device is also used to trigger the start of the exposure of each camera frame, using another one of its digital outputs. Moreover, it is also used to record signals from the point detectors, using its analog inputs. The synchronicity between the DAQ device's inputs and outputs is ensured by clocking with a single timer output. An Arduino Uno is used for controlling the 850 nm guide light, temperature and humidity sensor, and stepper motors (through stepper drivers).

\subsection{Software}
The instrument is controlled using code that is flexible enough to allow a wide range of measurement protocols to be run. It is written in a language (Python) that is relatively simple to grasp for beginners and is already widely used by researchers. Individual experiments are written and run using a single Python script. Example scripts for running each of the protocols used in the demonstration experiments detailed in Section~\ref{subsec:Illustrative Applications} are provided. These scripts can be easily modified to run new protocols. Importantly, to ensure reproducibility, the scripts integrate the Sacred framework \cite{klaus_greff-proc-scipy-2017}, which automatically logs the experiment code, collected data, metadata, and more into a MongoDB database (MongoDB Inc., New York, NY, US). This information can be easily visualized using a browser-based dashboard called Omniboard \cite{subramanian_vivekratnavelomniboard_2025}, which provides an intuitive interface for tracking past experiments, reviewing collected data, and maintaining an organized workflow. For data analysis, example scripts used to process the data collected during the demonstration experiments are also provided. Aside from code to run the main experiments, accessory codes are also provided: for measuring the illumination homogeneity using actinometry (Subsection~\ref{subsec:Validation of illumination homogeneity}); calibrating light intensity (Subsection~\ref{subsec:Calibration of irradiance}); performing harmonic correction of sinusoidal waveforms (Subsection~\ref{subsec:Sinusoidal light harmonics correction}); and controlling the sample platform's motors and 850 nm guide light. Section 2 of the Supporting Information contains full details on all scripts and the necessary device connections for each to run.

\section{Illumination Characterization, Calibration and Correction}     %%%%% CALIBRATION AND VALIDATION

\subsection{Measurement of Illumination Homogeneity} \label{subsec:Validation of illumination homogeneity}

Heterogeneity of intensity in the illuminated zone can impact experiments and the interpretation of their results, and is thus important to characterize. The optical simulation of the illumination system (Subsection~\ref{subsec:Optical Design}) provided a map of the light intensity which could be expected, yet the built system may deviate from this for many reasons and so it is important to verify it experimentally. As such, we demonstrate here one of the useful actinometry-based protocols detailed in~\cite{lahlou_fluorescence_2023} for the spatial measurement of light intensity. The protocol uses an RSFP called Dronpa-2, and relies on following its photoconversion from its bright to dark state, the rate of which being dependent on the light intensity, allowing its inference. Given that the simulation showed no significant wavelength-dependent differences in spatial distribution, the measurement was carried out using the 470~nm source only, corresponding to one of the two wavelengths where Dronpa-2 can be photoconverted. Other actinometers, as detailed in~\cite{lahlou_fluorescence_2023}, can be used where different wavelengths are to be tested.

\vspace{6pt}

\begin{figure}[t!]
\centering
\includegraphics[width=\textwidth]{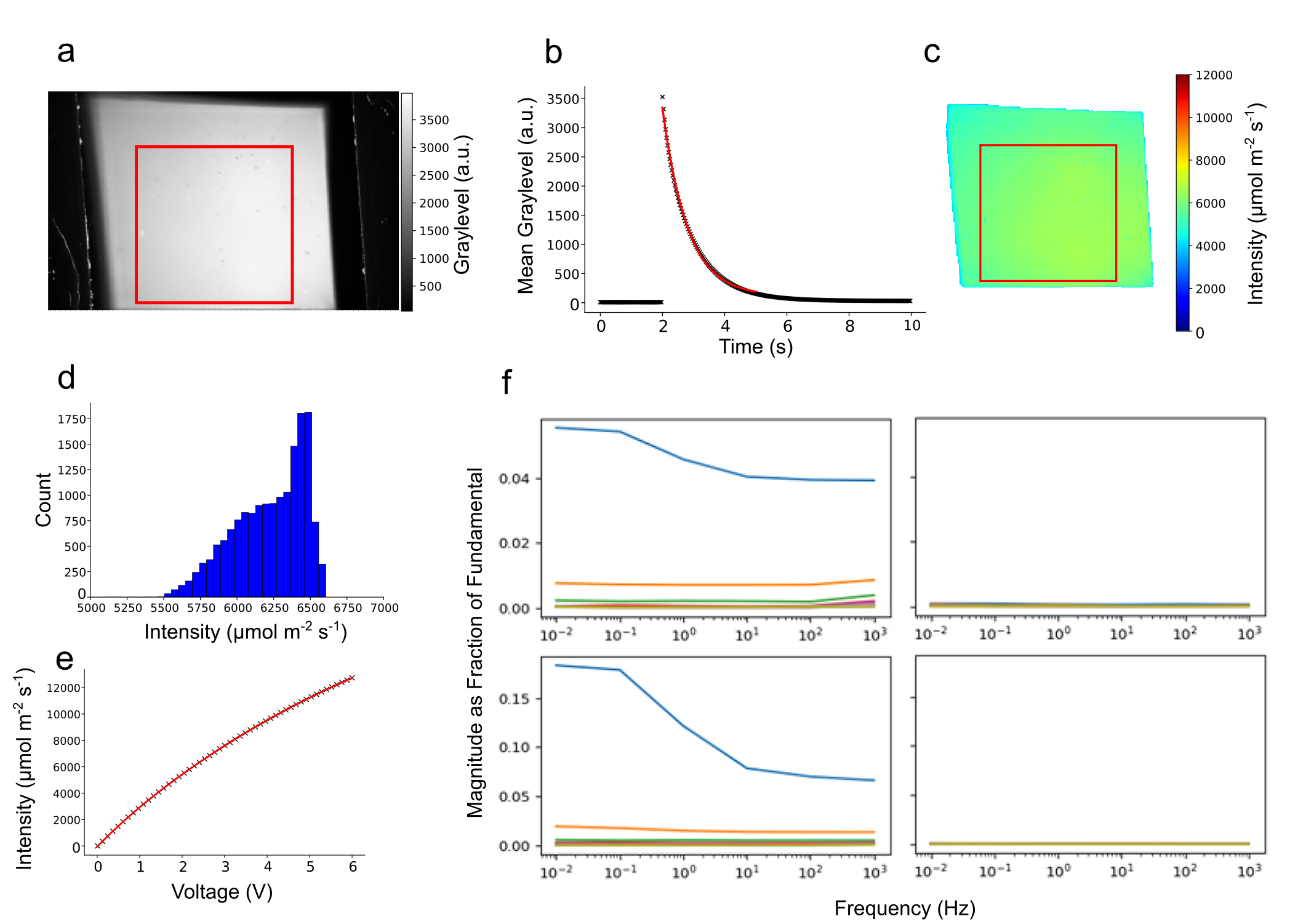}
\caption{\textbf{Characterization, calibration, and correction of the illumination system.} a-d: Characterization of the illumination's spatial homogeneity and intensity at a single wavelength. e: Calibration of light intensity as a function of LED level. f: Correction of harmonic distortion of sinusoidal waveforms caused by LED heating. (a) Representative frame from the fluorescence time series acquired during photoconversion of a 26 $\mu$M Dronpa-2 solution under 470 nm illumination. (b) Temporal evolution of mean fluorescence within the region-of-interest marked in (a), exhibiting a characteristic monoexponential decay; the extracted time constant enables inference of local intensity. (c) Spatial map of inferred intensity across the field-of-view, generated by pixel-wise fitting of exponential decay curves. (d) Histogram of pixel-wise intensity values within the illuminated zone, yielding a coefficient of variation of 3.9 \%, better than expected from the simulation. (e) Example calibration curve relating DAQ device output voltage to intensity, obtained using an optical power meter. (f) Harmonic content of sinusoidal oscillations of the 470 nm LED light source at different frequencies before (left) and after (right) application of the harmonic correction protocol, shown for both low (0.1–2.5 V, top row) and high (0.1–10 V, bottom row) voltage ranges. Harmonic magnitudes were quantified using fast Fourier transform (FFT) analysis. The iterative correction algorithm, which adds frequency-specific components in anti-phase to suppress nonlinear distortions, effectively eliminates higher-order harmonics while preserving the fundamental component. Legend: In panel (f), the second, third and fourth harmonics are represented in blue, orange, and green, respectively. Colors for higher-order harmonics are not specified, as their amplitudes are minimal and exhibit substantial overlap.}
\label{fig:FIG4}
\end{figure}

In order to perform the measurement, a 26 $\mu$M Dronpa-2 solution in PBS (Phosphate Buffered Saline, 0.01 M, [NaCl 0.138 M; KCl 0.0027 M], pH 7.4), held between two microscope cover slips separated by a 250~\textmu m-thick spacer, was suddenly illuminated with the 470 nm source at a constant intensity at the highest drive current level possible, inducing photoisomerization from the bright to dark state. This transition resulted in a monoexponential decrease in fluorescence, which was recorded at 30 fps using the camera and with the 540~nm emission filter. A single frame from this recording is shown in Figure~\ref{fig:FIG4}a, and the mean of the pixel values within the indicated region-of-interest (ROI), corresponding to $10 \times 10~\text{mm}^2$, for each frame plotted in Figure~\ref{fig:FIG4}b, and on a semi-log scale in Figure S101. In fitting a monoexponential function to such data, the time constant was determined to be \( 0.801\pm 0.010\,\text{s} \), and allowed for the mean light intensity to be inferred: 5700 $\mu\text{mol}\,\text{m}^{-2}\,\text{s}^{-1}$. In carrying out this fitting procedure for each pixel within the illuminated zone, a map of the tau values (shown together with a map of the uncertainty of their estimation in Figure S102) could be obtained, and then a map of light intensity (Figure~\ref{fig:FIG4}c), and the corresponding histogram of the intensity values within the ROI (Figure~\ref{fig:FIG4}d). The coefficient of variation, in percentage terms, was determined to be 3.9~\% - better than expected by simulation.

%%%%% FIGURE 4

\subsection{Calibration of Light Intensity} \label{subsec:Calibration of irradiance}
There are two core methods which can be used for calibration of intensity in this system. One can exploit the actinometry-based methods detailed in \cite{lahlou_fluorescence_2023} and \cite{lahlou_leaves_2024}, or an optical power meter can be used. In this section, given that one of the actinometry-based protocols was already explored in the previous subsection, measurement using a power meter is explored. One key disadvantage of using a power meter is that it assumes that light intensity is homogeneous. In this case, however, given the high degree of homogeneity of the illumination system, this issue is negligible. 

\vspace{6pt}

In order to measure intensity with the power meter, one must measure the optical power and also the area of the illuminated zone, the intensity being the optical power divided by the area. In order to measure optical power, the meter must be set to the wavelength to be measured, and its sensor placed in the appropriate side of the mount at the sample position. The placement for the left illumination arm is shown in Figure S64. This placement ensures the sensor is perpendicular to the illumination's optical axis, minimizing reflections from the sensor’s glass cover. Regarding the area, one way to determine it involves imaging the illuminated zone, measuring the area through image processing, and converting to units of $\text{m}^{2}$ using the pixel-to-m relationship of the imaging system. ImageJ \cite{schneider_nih_2012} can be a useful software program for this. The power value and area can then be used to determine the intensity in units of $\text{W}\,\text{m}^{-2}$. A Jupyter Notebook script is provided which automates this calculation, and also reports the intensity in units of $\mu\text{mol}\,\text{m}^{-2}\,\text{s}^{-1}$.

\vspace{6pt}

Since this is only pertinent for measurements at single intensity levels, another script is provided which automates this for a range of levels: it incrementally adjusts the LED level, records the power, and converts it into intensity. The result is a calibration curve of DAQ device output voltage versus intensity, which can be later used when programming experiments. An example of such a calibration curve is shown in Figure~\ref{fig:FIG4}e for the 470~nm source in the $5 \times 5~\text{mm}^2$ configuration. This automated script can only be used when the LED driver is being used to drive the LEDs.

\subsection{Sinusoidal Light Harmonic Correction} \label{subsec:Sinusoidal light harmonics correction}

When driving LEDs, a non-linear relationship between input current and output intensity is commonly observed, primarily due to heating effects. This non-linearity poses challenges when precise waveforms are required (e.g., for generating perfectly sinusoidal illumination). In practice, when a pure sinusoidal electrical signal is applied to the LEDs, the resulting optical waveform often contains harmonics. The extent of this distortion depends on the drive current level and the associated heating effects. In this subsection, we detail a protocol we developed to correct such distortions~\cite{Pellissier2022a}, and show results of its application to each of the LEDs in the system. It should be noted that at lower current levels, harmonic content is minimal and correction is generally unnecessary. For this reason, the harmonic correction protocol was applied only when using the LED driver, and not when driving LEDs directly with the waveform generator. 

\vspace{6pt}

To prepare the system for calibration, the camera is replaced with a point detector for signal collection. A mirror, held in a custom 3D-printed mount (shown in Figure S63), is placed at the sample position to direct light from the illumination arm to the detector. Since this setup directs a significant amount of light to the detector, the aperture of the camera objective is reduced to prevent saturation.

\vspace{6pt}

The harmonic content of the sinusoidally modulated illumination was assessed prior to applying harmonic correction. Measurements were conducted at DAQ device output voltage ranges (peak-to-trough) of 0.1–2.5 V and 0.1–10 V, and across modulation frequencies spanning 0.01–1000 Hz. Harmonic magnitudes, expressed relative to the magnitude at the fundamental frequency, for the 470 nm LED and covering the first 10 harmonics, extracted via Fast Fourier Transform (FFT) analysis, are shown in the left panels of Figure~\ref{fig:FIG4}f. To suppress unwanted harmonics, components at the harmonic frequencies are added to the original sinusoidal signal in anti-phase. The resulting waveform is then normalized to match the original signal’s upper and lower voltage bounds. This corrected signal is used to modulate the LEDs again, and the output is re-evaluated. If the residual harmonic content exceeds a defined threshold, an additional correction is applied. This process is repeated iteratively until sufficient suppression is achieved. The effectiveness of the correction is demonstrated by the substantial reduction in harmonic content visible in the right panels of Figure~\ref{fig:FIG4}f. Results for all LEDs are provided in Figures S87–S91.

\section{Illustrative Demonstrations} \label{subsec:Illustrative Applications}    %%%%% ILLUSTRATIVE APPLICATIONS

\subsection{Photosynthetic Species}

\subsubsection{Photosynthetic Parameters Extraction}\label{subsec:Photosynthetic parameters extraction}

In this experiment, we aimed to demonstrate the capability of the instrument to perform basic pulsed-light protocols. As such, we aimed to apply an adapted version of the protocol described in \cite{nedbal_kinetic_2000} which was developed to recover physiologically relevant parameters of plants. Here the intention was to show that such a type of experiment can be run, which requires very short pulse durations, precise timing and synchronization between light sources and the camera, but we cannot claim here that we were able to accurately retrieve such physiological parameters in their precise definitions, this is well beyond the scope of this paper. The physiological parameters in question are the minimum fluorescence, $F_0$, maximum fluorescence, $F_M$ and non-photochemical quenching, $NPQ$, whose definitions are given in \cite{nedbal_kinetic_2000}. Here, since we cannot be sure to measure exactly these, we will use the notation $\mathcal{F}_0$, $\mathcal{F_M}$ and $\mathcal{NPQ}$. The protocol we implemented was applied to a leaf very shortly after it was excised (within 15 minutes), and again after the leaf was kept in the dark for 3 hours, so that some differences could be seen. The illumination arms were reconfigured to have two 470 nm sources, one on each arm (Figure S34). One of the sources was used only for measuring pulses, while the second was used for saturating light pulses and actinic light. Throughout the entire protocol, the second source was briefly switched off during the measuring flashes, to maintain consistent excitation conditions for each frame acquired. Here, the 690~nm emission filter was selected.

\vspace{6pt}

First, the leaf was placed between two microscope slides, and then dark-adapted for 15 minutes. The measurement sequence began with illuminating the sample with short (10~$\mu\text{s}$) blue measuring pulses (470~nm source) at an intensity of 2000~$\mu\mathrm{mol}\,\mathrm{m}^{-2}\,\mathrm{s}^{-1}$, delivered at a frequency of 20~Hz for 3~s.  Images were captured simultaneously at the same frequency, using a 10~$\mu\text{s}$ exposure time synchronized to the excitation pulses. This allowed the determination of $\mathcal{F}_0$. Subsequently, a 2~s blue saturating pulse (470~nm source) at 2000~$\mu\mathrm{mol}\,\mathrm{m}^{-2}\,\mathrm{s}^{-1}$ was applied while the measuring pulses and synchronized imaging continued at 20~Hz, for the measurement of $\mathcal{F_M}$. After a 15~s dark adaptation period, the intensity of the second blue source was set to 300~$\mu\mathrm{mol}\,\mathrm{m}^{-2}\,\mathrm{s}^{-1}$, providing actinic light. The measuring flashes and synchronized imaging were maintained at 20~Hz for 15~s. Following this phase, a second 2~s pulse of high-intensity blue light (2000~$\mu\mathrm{mol}\,\mathrm{m}^{-2}\,\mathrm{s}^{-1}$) was applied, again with synchronized blue measuring flashes and imaging at 20~Hz. This allowed the determination of $\mathcal{F_M'}$, from which $\mathcal{NPQ}$ can be calculated.

\vspace{6pt}

\begin{figure}[htbp]
\centering
\includegraphics[width=12cm]{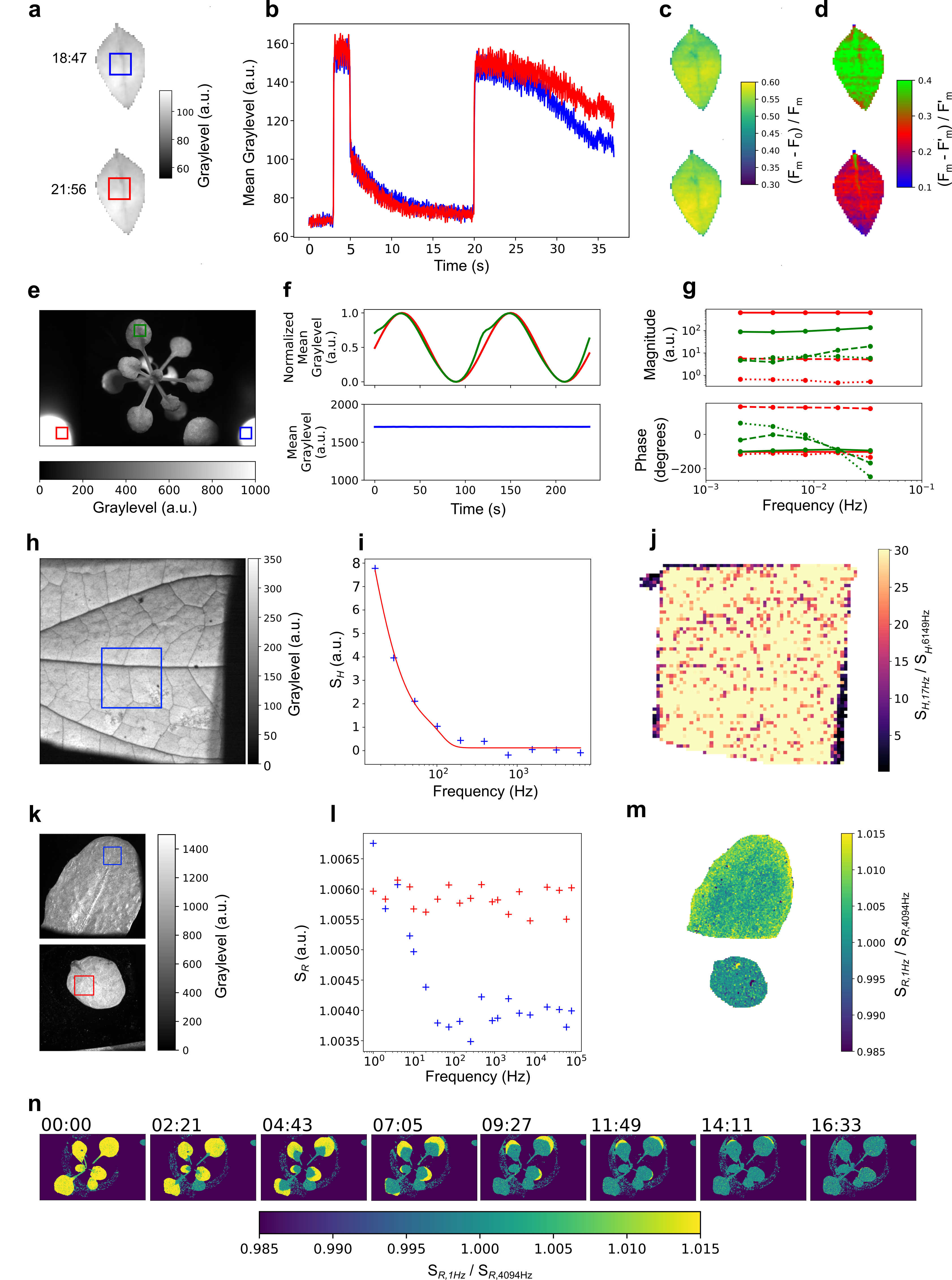}

\caption{
\textbf{Demonstration of the instrument’s capability for pulsed and modulated excitation fluorescence measurements in plants.}
(a–d) Pulsed light protocol: (a) Averaged fluorescence image with ROIs; (b) Mean fluorescence signals at time 0 and three hours post-excision (colors correspond to ROIs in a); (c–d) Pixelwise maps of $\mathcal{(F_M - F}_0\mathcal{)/F_M}$ and $\mathcal{NPQ}$ showing spatial and temporal variations.
(e–g) Frequency response protocol: (e) Fluorescence image with ROIs [red: modulated LED reference, green: sample, blue: camera heating reference]; (f) Time-resolved signals from leaf and reference fibers [colors as in e]; (g) Frequency response from unnormalized signals at the first three harmonics [colors as in e; solid, dashed, and dotted lines denote harmonics 1, 2, and 3].
(h-j) HIOM protocol: (h) ROI selection on single frame; (i) HIOM observable versus frequency [with eye guideline]; (j) Pixelwise map of HIOM observable ratios. 
(k-m) RIOM protocol: (k) ROI selections on single frames; (l) RIOM observable versus frequency for the untreated and DCMU-treated leaves; (m) Pixelwise RIOM ratio maps. 
(n) Tracking of DCMU uptake in an intact \textit{Arabidopsis thaliana} plant using RIOM over 16+ hours.
}

\label{fig:FIG5}
\end{figure}

Figure~\ref{fig:FIG5}a shows averaged images of the leaf captured at each probe time point, along with regions of interest (ROIs) used for signal extraction. The corresponding mean signals are presented in Figure~\ref{fig:FIG5}b, illustrating clear differences between the two time points. Pixelwise calculations of $\mathcal{(F_M - F}_0\mathcal{)/F_M}$ and $\mathcal{NPQ}$ were performed to obtain spatial maps, shown in Figures~\ref{fig:FIG5}c and \ref{fig:FIG5}d. These maps reveal local variations in fluorescence parameters and indicate a decrease in NPQ between the two time points.

\subsubsection{Frequency Response}
In this section, we demonstrate the use of the instrument for measuring the fluorescence response of a plant in the frequency domain, a powerful alternative approach for interrogating photosynthetic species. It has been used previously for examining the operation frequency range of photoprotective regulatory mechanisms for example \cite{niu_plants_2023}. To this end, after 15 minutes of light acclimation at the average intensity of the sinusoidal excitation, the system was sequentially excited with sinusoidally modulated 470~nm light with periods ranging from 480 to 30~s, average intensity of 450~$\mu\mathrm{mol}\,\mathrm{m}^{-2}\,\mathrm{s}^{-1}$, and a modulation amplitude of 350~$\mu\mathrm{mol}\,\mathrm{m}^{-2}\,\mathrm{s}^{-1}$, while fluorescence image frames (with the 690~nm emission filter selected) were continuously recorded at 40 frames per period. The corresponding fluorescence responses were analyzed using Fourier analysis.

\vspace{6pt}

Figure~\ref{fig:FIG5}e shows a single fluorescence image frame acquired during the experiment. In Figure~\ref{fig:FIG5}f, two periods of the time-resolved fluorescence signals at a selected modulation frequency are presented, obtained by averaging over ROIs corresponding to: (1) the leaf surface, (2) the illumination reference fiber, and (3) a constant-intensity LED reference fiber. This highlights the advantage of having reference signals directly embedded within the imaging field, allowing direct and straightforward visualization of deviations of the fluorescence from the excitation waveform. The constant LED reference also confirms the absence of intensity drifts attributable to camera heating. Figure~\ref{fig:FIG5}g shows the frequency response within the leaf ROI at the fundamental frequency (solid green), as well as the second (dashed green) and third (dotted green) harmonics. The corresponding responses from the excitation reference signal are shown in red.

\subsubsection{RIOM and HIOM}
We demonstrate here the use of the instrument for the application of two newly developed protocols, \textit{Rectified Imaging under Optical Modulation} (RIOM) and \textit{Heterodyne Imaging under Optical Modulation} (HIOM), which enable the imaging of fast photochemical events using a standard low-cost low-frequency camera\cite{merceron_periodic_nodate}. Both approaches involve probing the sample across a range of modulation frequencies and recording an observable at each frequency, though the observables differ between the two protocols. In the first approach, RIOM, the observable is the mean signal averaged over all periods at each modulation frequency. In contrast, HIOM uses two simultaneously modulated light sources, with one modulated at a fixed frequency offset (here, 1~Hz) from the other. The observable is the magnitude of the out-of-phase component at the offset frequency. When a luminophore’s behavior follows a two-state model, the fitting functions described in~\cite{merceron_periodic_nodate} can be applied to the frequency dependence of the RIOM and HIOM observables to extract the characteristic photoactivation time. When the luminophore deviates from a two-state model and fitting is not feasible, the frequency response can instead serve as a kinetic fingerprint, still containing valuable dynamic information. In this section, we apply these protocols to leaves.

\vspace{6pt}

We first demonstrate the application of the HIOM technique to a leaf. Here two identical light sources were required on each illumination arm. The setup of the illumination arms was modified such that two 470~nm sources could be piloted simultaneously. The setup is the same as that detailed for the PAM-like experiment (Subsubsection \ref{subsec:Photosynthetic parameters extraction}). Sinusoidally modulated 470~nm illumination was applied, with modulation frequencies spanning a broad range relevant to plant fluorescence induction kinetics: [17~Hz; 6149~Hz]. A second light source was modulated at a frequency 1 Hz lower than the primary illumination. The mean irradiance of each light source was set to 100~$\mu\mathrm{mol}\,\mathrm{m}^{-2}\,\mathrm{s}^{-1}$, with a modulation amplitude of 100\%. The camera exposure was set to 80~ms, with a frame rate of 12 fps, and the 690~nm emission filter selected.

\vspace{6pt}

Figure~\ref{fig:FIG5}i shows the HIOM observable (the out-of-phase magnitude at 1~Hz) as a function of excitation frequency for the mean frame data within the region of interest (ROI) indicated in Figure~\ref{fig:FIG5}h. This data represents the kinetic response associated with the reversible photoactivation of the photosynthetic system. The plot demonstrates a decline toward higher frequencies, followed by a stabilization. Given the complexity of chlorophyll fluorescence dynamics, more intricate than the two-state model on which RIOM and HIOM were developed, a direct fit was not performed on the data. Instead, a ratio between the HIOM observable at 17~Hz and 6149~Hz was calculated for each pixel, forming a spatial map of the difference in response at these two frequencies (Figure~\ref{fig:FIG5}j).

\vspace{6pt}

For RIOM, sinusoidally modulated 470~nm illumination was applied with modulation frequencies spanning [1~Hz; 80,000~Hz] to both an untreated leaf and another leaf after being immersed in a solution of 3-(3,4-dichlorophenyl)-1,1-dimethylurea (DCMU), a herbicide that inhibits photosynthesis, for several hours. At each modulation frequency, the camera was operated at 0.5 fps with a 1-second exposure to integrate an integer number of oscillations.

\vspace{6pt}

Figure~\ref{fig:FIG5}l shows the ratio of the average signal level during modulation to that under constant (unmodulated) illumination, plotted as a function of modulation frequency for the two leaves. A ratio was calculated at each frequency to correct for drift in the constant-light fluorescence signal, which is not steady over the duration of the protocol. These values were derived from the mean signal over the pixels within the ROIs indicated in Figure~\ref{fig:FIG5}k. The results reveal clear differences in the kinetic responses of the untreated and DCMU-treated leaves. These differences are further illustrated in the spatial maps in Figure~\ref{fig:FIG5}m, which show the RIOM observable ratio at 1~Hz versus 4094~Hz for each pixel. Further interpretation of these findings is beyond the scope of this manuscript.

\vspace{6pt}

In order to demonstrate the versatility of the instrument further, an experiment was performed in which DCMU was applied to the roots of a small \textit{Arabidopsis thaliana} (Col-0) plant held in a plant pot. During the uptake of the herbicide, the RIOM protocol was applied at intervals over the course of 16+ hours. Figure~\ref{fig:FIG5}n demonstrates the power of this technique in tracking the uptake of DCMU.

\subsection{Speed-OPIOM and RIOM Applied to Reversibly Photoswitchable Proteins in Droplets}

We demonstrate here the capability of the instrument to implement the Speed-OPIOM\cite{querard_resonant_2017} and RIOM\cite{merceron_periodic_nodate} methodologies on droplets of RSFP solutions. Speed-OPIOM leverages the distinct photoswitching kinetics of various RSFPs to enable selective imaging. By optimizing the out-of-phase amplitude through specific intensities and frequencies of periodic excitation, a particular RSFP can be selectively detected using phase-sensitive detection.

\begin{figure}[t!]
\centering
\includegraphics[width=\textwidth]{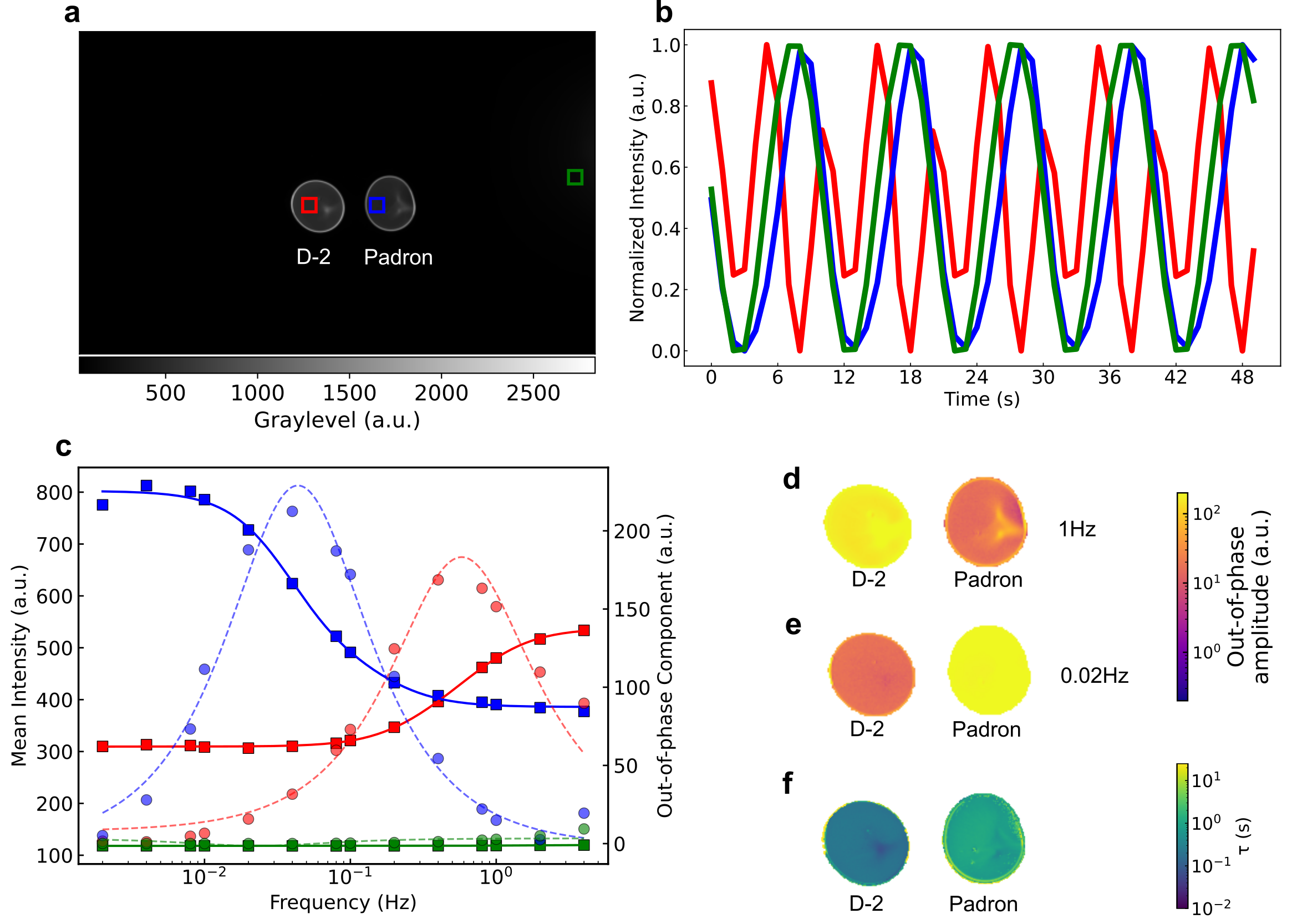}
\caption{
    \textbf{Dynamic fluorescence response and analysis of Dronpa-2 and Padron RSFPs under sinusoidal modulation.} (a) Mean intensity across the experimental dataset, illustrating the steady-state fluorescence of both Dronpa-2 and Padron. Due to overlapping emission spectra, the proteins cannot be distinguished based solely on their steady-state fluorescence intensities. (b) Fluorescence signals averaged over the ROIs highlighted in (a), collected at a modulation frequency of 0.04 Hz, showcasing the differing dynamic responses of Dronpa-2 and Padron relative to the inert reference region. (c) Frequency-dependent behavior of the Speed-OPIOM and RIOM observables across all modulation frequencies, with distinct profiles for Dronpa-2 and Padron. The reference ROI exhibits negligible response. (d, e) Contrast-enhanced images obtained by calculating the Speed-OPIOM observable on a pixel-by-pixel basis, with modulation frequencies of 0.02 Hz (d) for Padron and 1 Hz (e) for Dronpa-2. (f) Spatial map of photoswitching time constants derived from pixelwise fitting of the RIOM observable, providing another mechanism for differentiating the two RSFP species based on their photoswitching kinetics. Colors in (b,c) correspond to ROI outline colors in (a).
}
\label{fig:FIG6}
\end{figure}

\vspace{6pt}

Individual droplets of Dronpa-2 and Padron solutions were confined between two microscope coverslips separated by a 0.25 mm spacer (Gene Frames AB0578; Thermo Scientific Inc., Waltham, MA, USA). Details of the kinetic characteristics of these proteins can be found in \cite{querard_resonant_2017}. It is noteworthy that, although the fluorescence emission spectra of Dronpa-2 and Padron largely overlap, their photoisomerization cross sections differ significantly, resulting in distinct photoswitching kinetics under identical illumination conditions. As shown in Figure~\ref{fig:FIG6}a, which presents the mean of all collected frames, the two proteins cannot be distinguished based solely on their steady-state fluorescence intensities.

\vspace{6pt}

The experimental protocol remained identical for the extraction of both the Speed-OPIOM and RIOM observables, with differences arising solely in the downstream processing. Samples were subjected to sinusoidally modulated illumination at 470 and 405 nm, oscillating in antiphase over a frequency range spanning 0.002 to 4 Hz. The average intensity of the 470 nm excitation was maintained at 12000 $\mu \mathrm{mol} \, \mathrm{m}^{-2} \, \mathrm{s}^{-1}$, while that of the 405 nm light was set to 6000 $\mu \mathrm{mol} \, \mathrm{m}^{-2} \, \mathrm{s}^{-1}$, both at 100\% modulation depth. Removal of dichroic mirrors and mounting of one source per illumination arm, akin to that done for the PAM-like experiment (Subsubsection \ref{subsec:Photosynthetic parameters extraction}), as well as reducing the illumination area to 5 × 5 mm$^2$, was required to achieve these high light intensities. Prior to data acquisition, the harmonic correction protocol was applied to ensure highly pure sinusoidal illumination. Imaging was conducted at a frame rate corresponding to 15 frames per period, with an exposure time of 50 ms, and the 540~nm emission filter selected.

\vspace{6pt}

Fluorescence signals averaged over the ROIs marked in Figure~\ref{fig:FIG6}a, collected at 0.04 Hz, illustrate the differing dynamic responses of Dronpa-2 and Padron relative to the inert reference region. Following extraction of the Speed-OPIOM and RIOM observables across all modulation frequencies, the frequency-dependent behavior displayed in Figure~\ref{fig:FIG6}c was obtained. As expected, the reference ROI exhibited a negligible response. In contrast, Dronpa-2 and Padron demonstrated distinct RIOM profiles, with the characteristic cutoff frequency for Dronpa-2 appearing at a higher frequency relative to Padron, consistent with its faster switching kinetics. Fitting of the RIOM curves, using the relevant equation described in \cite{merceron_periodic_nodate}, yielded characteristic switching time constants of 0.22 s and 1 s for Dronpa-2 and Padron, respectively. Fitting was also performed on the out-of-phase Speed-OPIOM signals, serving as a visual aid to highlight the correspondence between the RIOM cutoff positions and the Speed-OPIOM peak responses.

\vspace{6pt}

To generate contrast-enhanced images, the Speed-OPIOM observable was calculated on a pixel-by-pixel basis. By selecting the modulation frequency appropriately, 0.02 Hz to enhance Padron (Figure~\ref{fig:FIG6}d) and 1 Hz to enhance Dronpa-2 (Figure~\ref{fig:FIG6}e), protein-specific contrast images were obtained. Finally, pixelwise fitting of the RIOM observable provided a spatial map of the photoswitching time constants (Figure~\ref{fig:FIG6}f), offering an additional way to differentiate between the two RSFP species.

\subsection{Electroluminescence of Optoelectronic Devices} %%%%% Semiconductor Electroluminescence

In this section, we further illustrate the versatility of the system by demonstrating its applicability to scenarios where the sample is excited electrically, as opposed to via optical illumination. Specifically, we present a preliminary experiment employing a protocol akin to RIOM, albeit with electrical power modulation, on an organic photovoltaic (OPV) cell and a blue LED (L1RX-BLU1000000000, Lumileds, San Jose, CA).

\vspace{6pt}

In general, the ultimate performance of OPV devices is limited by electronic processes occurring at different levels, which include geminate recombination of photo-generated electron-hole pairs, trap-assisted recombination, trap-limited transport of charge carriers, and interfacial recombination. Capacitive effects also largely govern the dynamic response of a device under modulated excitation, which is particularly pertinent in optical communication applications, for example~\cite{zhang_organic_2015}. Probing these mechanisms can be done using dedicated characterization techniques, including time-resolved photoluminescence or electroluminescence imaging, which have been shown to be highly versatile~\cite{bercegol_time-resolved_2018}. Indeed, a good OPV should in general be a good emitter as well, and imaging techniques have been recently developed and adapted to allow discrimination between factors limiting device performance. Here we illustrate the relevance of our instrument for imaging the dynamic emission of a reference laboratory-scale OPV device upon modulated electrical excitation. The OPV device is based on a classical inverted device architecture (transparent electrode/electron transport layer/active layer/hole transport layer/metallic top electrode), using the benchmark PM6:Y6 active layer blend (see~\cite{shoaee_what_2024}).

\vspace{6pt}

\begin{figure}[b!]
\centering
\includegraphics[width=\textwidth]{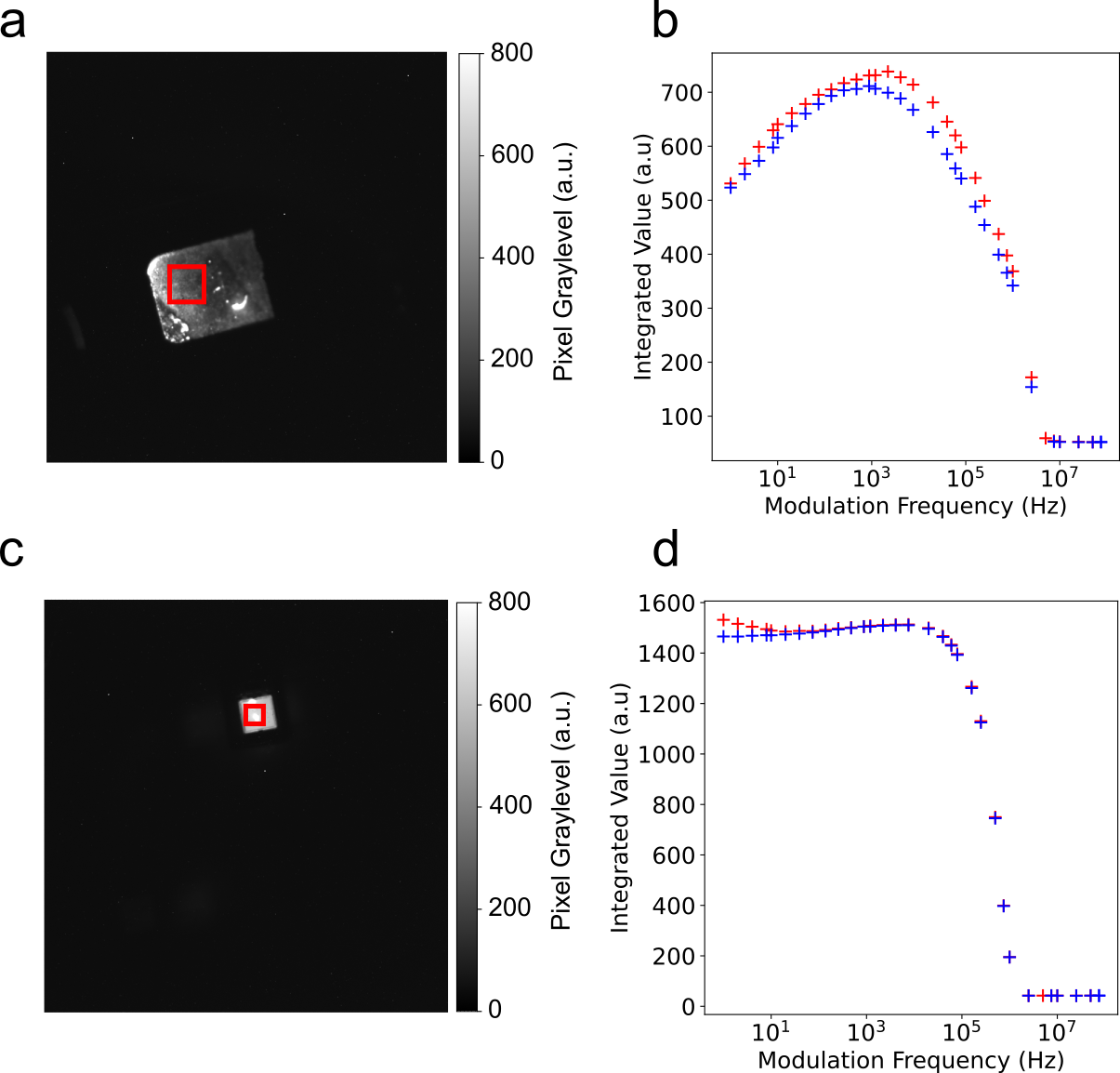}
\caption{\textbf{Response of OPV cells and LEDs under periodic electrical excitation.} (a) and (c) show the mean images of the OPV cell and LED, respectively, with the ROIs used for signal analysis marked. (b) and (d) display the corresponding RIOM observable as a function of frequency. The devices were sequentially subjected to square wave-modulated electrical excitation (0 to 10 V) at frequencies from 1 Hz to 75 MHz (red) and then 75 MHz to 1 Hz (blue), with a 1-second exposure time and no emission filter.}
\label{fig:FIG7}
\end{figure}

A square wave-modulated electrical signal, ranging from 0 to 10 V, was applied over a frequency range extending from 1 Hz to 75 MHz using the waveform generator. Simultaneously, image frames were captured with exposure times of 1 second, and no emission filter employed. The frequency range was probed sequentially from low to high and then high to low to detect any hysteresis. The resulting RIOM observable as a function of frequency is depicted in Figure~\ref{fig:FIG7}b, derived by integrating the data within the ROI marked in Figure~\ref{fig:FIG7}a, which shows the first frame captured under 1 Hz oscillation. 

\vspace{6pt}

Interestingly, the frequency-evolution of the integrated signal peaks for a modulated frequency of approximately 1kHz, before a clear cut-off is observed around 100-500 kHz. An in-depth analysis and interpretation is beyond the scope of this paper, yet it can be said that the response clearly evidences the limitation of the device response due to its geometric capacitance. Further work is required to evaluate whether the measurement can bring more insight on the dynamic behavior of the OPV device. Specific investigations using modulated illumination could also be easily conducted using the instrument, exciting the constitutive layers of the OPV to shed light on intrinsic electronic processes (e.g., recombination and transport) and aging effects.

\vspace{6pt}

A parallel experiment was conducted with the LED, yielding the response curve shown in Figure~\ref{fig:FIG7}d, based on the data from the ROI marked in Figure~\ref{fig:FIG7}c. Although the creation of response images and the detailed interpretation of this data falls outside the scope of this manuscript, it is evident that the system exhibits a frequency-dependent response, which may provide valuable insights into its spatial and transient behavior.

\vspace{6pt}

In both cases, it was shown that the instrument can perform protocols capable of revealing potentially useful information about OPV and LED devices. Here, the dynamic response of each is obtained, which is mostly limited by the geometric capacitance in this case. The fact that different cut-off frequencies were seen is in line with what would be expected given their respective active areas. The OPV, having a larger active area ($0.2~\text{cm}^2$), shows a cut-off at lower frequency than the LED, which has a smaller active area ($0.01~\text{cm}^2$). Such a protocol could prove useful in the quality control of optoelectronic devices due to the possibility to quickly image the response of the whole active area. The flexibility of the instrument makes it straightforward to modify such protocols in order to optimize them, or to create novel ones.

\section{Conclusion}     %%%%% CONCLUSION
In this work, we have presented a versatile luminescence macroscope system; detailed its capabilities and design; performed characterization, calibration and correction protocols; and provided representative demonstrations of its capabilities. To facilitate adoption and adaptation, we have made available comprehensive build instructions and design files. Owing to its flexibility with respect to illumination and imaging sequences, excitation and emission wavelengths, and supported sample geometries, the system is well-suited to a wide range of experimental paradigms beyond those illustrated here. We hope that, by providing a fully open and accessible design, this platform may help to lower the barrier for research teams who may not have the resources to undertake bespoke instrumental development or the complex modification of commercial systems. We anticipate that this instrument could offer a useful basis for future innovations in imaging protocols.

\section*{Funding}     %%%%% FUNDING
This work was supported by the EIC Pathfinder Open project DREAM (Grant Agreement number 101046451).

\section*{Acknowledgments}     %%%%% ACKNOWLEDGMENTS
Mhairi L. H. Davidson (CNRS – Sorbonne Université, UMR7622, Paris, France) is acknowledged for her assistance in the cultivation of Arabidopsis thaliana. We also thank Yuriy Shpinov and Melike Bekçi (PASTEUR, Department of Chemistry, École Normale Supérieure, PSL University, Sorbonne University, CNRS, Paris, France) for providing access to fluorescent beads.

\section*{Disclosures}     %%%%% DISCLOSURES
The authors declare no conflicts of interest.

\section*{Data Availability}     %%%%% DATA AVAILABILITY
All data and analysis codes associated with this manuscript are available at the following DOI: 10.5281/zenodo.15632946.

\section*{Supporting Information}     %%%%% SUPPORTING INFORMATION
See Supplement 1 for supporting content.

%%%%%%%%%%%%%%%%%%%%%%% References %%%%%%%%%%%%%%%%%%%%%%%%%

%%%%%%%%%% If using BibTeX:
\bibliography{sample}

\end{document}